\documentclass[conference]{IEEEtran}

\usepackage[T1]{fontenc}
\usepackage{graphicx}
\usepackage{amsmath,amssymb}
\usepackage{cite}
\usepackage[utf8]{inputenc}
\usepackage[hidelinks]{hyperref}
\usepackage{multirow}
\usepackage[protrusion=true,expansion=false]{microtype}

\begin{document}

\title{An Empirical Evaluation of Prompt Injection Vulnerabilities in Large Language Models Across Multilingual and Obfuscated Attack Scenarios}
 \author{
 Çağlar Uysal,
 Baturay Birinci,
 Süha Orhun Mutluergil,
 Orçun Çetin\\
 Sabancı University, Turkey\\
 \{caglaruysal, baturaybirinci, suha.mutluergil, orcun.cetin\}@sabanciuniv.edu
 }

\maketitle

%
\begin{abstract}
Large Language Models (LLMs) have rapidly evolved, transforming industries by automating complex tasks and generating human-like content. However, as their adoption accelerates, prompt injection vulnerabilities have become increasingly apparent. Malicious actors exploit these weaknesses to generate phishing emails, deceptive websites, and malware, posing serious security risks. This paper presents an empirical evaluation of six state-of-the-art LLMs (DeepSeek, GPT, Gemini, Grok, Llama, and Qwen) under diverse adversarial prompt scenarios, including direct and multi-stage obfuscated attacks across multiple languages and character encodings. The proposed framework measures how effectively current LLMs resist manipulation into performing harmful actions.

Our findings reveal systematic vulnerabilities across all tested models. Even direct prompt injections frequently induce the generation of phishing content, websites, and malware, while elaborate prompts achieve even higher malicious compliance rates, particularly for phishing. Models such as DeepSeek, Gemini, and Grok show especially high susceptibility under complex instructions. Notably, non-English languages consistently exhibit higher compliance rates than English, exposing significant gaps in multilingual safety alignment. Although simple character encodings reduce malicious outputs, they do not eliminate them. These results highlight persistent challenges in LLM safety and underscore the urgent need for stronger defenses and improved security mechanisms to support the ethical and secure deployment of LLMs in cybersecurity sensitive contexts.

\end{abstract}


%
\section{Introduction}
\label{sec:introduction}

Generative Artificial Intelligence (GenAI) continues to transform society through automation, productivity enhancement, and innovation across diverse domains. Large Language Models (LLMs), a prominent class of generative AI systems, have demonstrated remarkable capabilities in generating human-like text, assisting with decision-making, and automating complex tasks. In software development, LLM-based tools such as GitHub Copilot and ChatGPT have become integral to modern workflows, suggesting code snippets, debugging, and accelerating development cycles. 
However, the rapid expansion of LLM capabilities and their global deployment have elevated cybersecurity implications to a critical concern. In particular, these models are increasingly exploited through adversarial prompt manipulation, enabling the large-scale generation of phishing content, deceptive websites, and malicious code, thereby lowering the barrier to entry for cybercrime and amplifying the potential impact of automated attacks.

LLMs present significant security risks that have been systematically documented by initiatives such as the OWASP Top 10 for LLM Applications\footnote{OWASP Foundation, ``OWASP Top 10 for Large Language Model Applications.'' [Online]. Available: \url{https://genai.owasp.org/llm-top-10/}}, which identifies prompt injection as the most critical vulnerability. Prompt injection attacks occur when adversaries craft malicious inputs that manipulate LLMs into generating harmful or unintended outputs, bypassing safety mechanisms. The practical impact of such vulnerabilities has been demonstrated in real-world deployments. For example, in December 2023, a Chevrolet dealership’s ChatGPT-powered customer service chatbot was manipulated via prompt injection to appear to agree to sell a vehicle for \$1 \cite{chevy_prompt_injection_2023}. Although the interaction had no legal or transactional validity, it highlighted how inadequately constrained LLM-based systems can be coerced into producing misleading or policy-violating responses.

Existing safety mechanisms attempt to mitigate this risk through alignment training, safety policies, and prompt filtering. However, most publicly documented evaluations focus on English prompts and relatively simple, directly malicious requests. In practice, motivated adversaries are free to vary both the linguistic form and the technical encoding of their prompts. For instance, adversaries may request the same payload in Turkish or Russian, or conceal the malicious component of the instruction using character encodings such as Base64, ROT13, or hexadecimal in an attempt to bypass superficial filtering mechanisms. Today, there is remarkably little systematic evidence on how such choices of language and encoding affect the likelihood that a model will comply with malicious requests. This gap is particularly concerning given the global deployment of LLMs and the increasing availability of locally deployable models that operate outside tightly controlled platforms.

This study presents a systematic evaluation framework that assesses LLM compliance with malicious prompts across diverse scenarios, languages, and character encodings. We designed prompts targeting three categories of malicious artifacts: phishing emails, phishing webpages, and keylogger malware. The prompts are categorized into \emph{direct} and \emph{elaborate} groups. Direct prompts explicitly request malicious artifacts, such as phishing email text or malware source code. In contrast, elaborate prompts embed malicious requests within ostensibly legitimate scenarios (e.g., security awareness training, red team assessments) while carefully avoiding terminology that might trigger safety filters. Our evaluation includes 7 direct and 6 elaborate prompts for phishing emails, 7 direct and 6 elaborate prompts for phishing webpages, and 6 direct and 5 elaborate prompts for keylogger generation. This study evaluates the security behavior of six state-of-the-art large language models : DeepSeek, GPT, Gemini, Grok, Llama, and Qwen. Each unique prompt–language–model combination is tested over ten iterations to account for non-deterministic model behavior. Additionally, we investigate how different natural languages (English, Turkish, Russian, Simplified Chinese) and simple character encodings (Base64, ROT13, Hex) affect the likelihood that models will comply with malicious requests.

The analysis is conducted on a total of 15540 evaluated model responses, and the findings demonstrate that malicious request fulfillment is not an isolated edge case but a systematic phenomenon. Across all experimental conditions, 68.76\% of requests result in full malicious compliance, while an additional 12.08\% produce partial or stalling assistance, wherein the model supports the structuring or refinement of the attack without directly generating the final malicious payload. Taken together, 80.84\% of malicious requests are at least partially fulfilled. Substantial variation is observed across models, languages, and categories. Some models exceed 80\% full compliance, while even the most restrictive model still fulfills nearly three quarters of malicious requests when stalling assistance is included. Non-English languages consistently exhibit higher compliance than English, and elaborate, contextually rich prompts are markedly more successful than direct, blunt requests.

\subsection{Contributions}

The main contributions of this study are summarized below:

\begin{itemize}
  \item Our findings indicate that even direct prompt injections can successfully target all the LLMs in our study to generate phishing emails, websites, and malware. Moreover, high malicious request compliance rates of elaborate malicious prompts underscore a significant risk of misuse, emphasizing the need for developing sophisticated detection methods and security protocols tailored to address complex threats effectively. Many models show greater malicious compliance to elaborate malicious prompts, with this issue being particularly pronounced in scenarios related to phishing emails and websites.
  \item The study finds that LLMs such as Deepseek, Gemini, and Grok are highly effective at generating convincing malicious content, especially when given elaborate prompts.
  \item Interestingly, certain models demonstrate consistently high compliance rates with malicious prompts across nearly all categories (phishing emails, phishing websites, and malware). Of all the LLMs tested, some appear most inclined to fulfill harmful instructions, making them particularly concerning if misused.
  \item Lastly, non-English languages consistently exhibit higher compliance than English, while simple character encodings reduce but do not prevent malicious outputs, underscoring the critical need for multilingual and encoding aware safety mechanisms.
\end{itemize}

This paper proceeds with the following structure. Section~\ref{sec:related_work} outlines relevant prior work on LLM‑enabled cybercrime and prompt‑based safety evaluations. Section~\ref{sec:methodology} describes our threat model, prompt design, model selection, and experimental framework. Section~\ref{sec:results} presents our empirical findings across models, languages, encodings, and categories. Section~\ref{sec:discussion} discusses the broader security implications of these results and highlights key vulnerabilities. Section~\ref{sec:conclusion} concludes with recommendations and directions for future research on building more robust and globally safe LLMs.

\section{Related Work}
\label{sec:related_work}

This section reviews research that motivates our study and positions our contribution within prior work on LLM misuse, security threats, and safety evaluation. We group the literature into four themes: LLM-enabled phishing, LLM-assisted malware, prompt injection and alignment defenses, and empirical safety benchmarks.

\subsection{LLM-Enabled Phishing and Social Engineering}

This subsection summarizes evidence that LLMs can support social-engineering workflows, while highlighting gaps in empirical measurement of their real-world phishing effectiveness.
Recent studies suggest that LLMs can reproduce many linguistic and psychological patterns used in phishing, even when evaluated primarily in analytical settings. For example, using ChatGPT for semantic analysis of phishing emails shows that LLMs can identify technical and linguistic irregularities in ransomware-related campaigns \cite{fujima2023semantic}, reflecting attacker tactics such as urgency cues, authority mimicry, and obfuscation. Complementary work on phishing psychology emphasizes that successful messages exploit cognitive biases and behavioral triggers \cite{10_48009_2_iis_2023_107}, indicating why fluent, adaptive generation could meaningfully increase attack scalability.

However, systematic benchmarking of realism and success rates for LLM-generated phishing compared to human-crafted content remains limited. Work combining sentiment analysis with ChatGPT implies that LLM outputs could be tuned for emotional manipulation \cite{10_1109_sera61261_2024_10685564}, and broader analysis of LLM behavior supports their ability to simulate socially plausible communication \cite{10_1515_9781501518911_009}. In contrast, much of the tooling literature focuses on detection (e.g., APOLLO using GPT-4 for phishing detection) rather than evaluating LLMs as automated phishing agents \cite{10_1145_3733049}. Our work addresses two underexplored dimensions: multilingual phishing scenarios and encoded prompts intended to bypass safeguards, operationalized via \emph{malicious compliance rates}. 

\subsection{LLMs for Malware and Keylogger Generation}

This subsection reviews research connecting automation trends in malware development with concerns about LLM-assisted code generation and evasion.
While direct demonstrations of LLMs generating end-to-end malware remain relatively sparse, multiple studies describe trends that make AI-assisted malware development plausible. Ransomware research documents increasingly sophisticated, multi-stage monetization workflows \cite{10_5772_intechopen_108433}, and related discussions highlight how automation can lower barriers for less skilled adversaries \cite{10_22214_ijraset_2022_39787}. These patterns suggest that LLM-based assistance may accelerate malicious development by reducing iteration costs and enabling rapid adaptation.

Evasion-oriented work further motivates concern that AI assistance could amplify stealth and resilience. The ``digital camouflage'' framing highlights how adversarial methods can obscure malicious code and complicate detection \cite{boke2025digital}, while research on mobile malware manipulating LLM-based applications expands the threat surface to LLM ecosystems themselves \cite{huang2024strengthening}. Foundational analysis of malware circumvention and keylogger-related techniques shows how attackers adapt against defenses over time \cite{10_1109_malware_2013_6703691}, yet controlled experiments quantifying LLM-generated malware success against modern defenses remain limited. Our contribution complements this literature by reporting cross-model statistics on keylogger prompt compliance, shifting discussion from assumed risk to measurable behavior.

\subsection{Prompt Injection, Safety Alignment, and OWASP LLM Top 10}

This subsection reviews prompt injection as a dominant LLM security risk and summarizes mitigation approaches motivated by OWASP-style vulnerability reasoning.
Prompt injection has emerged as a core vulnerability in LLM-integrated systems, and OWASP-style frameworks provide a useful conceptual baseline. The OWASP Top 10:2021 taxonomy and its risk categories (e.g., injection, insecure design) offer a foundation for reasoning about LLM threats \cite{10_7717_peerj_cs_2821_table_10}, while efforts mapping OWASP Top 10:2021 to CWE Top 25:2023 illustrate how traditional vulnerability thinking carries into LLM contexts \cite{10_7717_peerj_cs_2821_table_10}. Forward-looking security discussions also anticipate increased emphasis on hardening LLM interfaces against adversarial prompting \cite{10_1109_icict64420_2025_11004742}.

Research on prompt injection emphasizes both attack mechanisms and defenses. Detection work on LLM-integrated applications describes how crafted prompts can override intended behavior \cite{10_53941_ijndi_2025_100013}, 
and injection-style demonstrations show how structured prompting can emulate traditional exploitation patterns \cite{10_7717_peerj_cs_2374_fig_9}. Proposed defenses include cryptographic input validation such as Signed-prompt \cite{Signed-prompt} and dynamic moving target defenses \cite{panterino2024dynamic}. Our study complements these efforts by empirically measuring harmful prompt success across models, languages, and encodings, providing evidence relevant to alignment design.

Prior studies have employed malicious prompt datasets, such as AdvBench and HarmBench, to evaluate the safety and robustness of LLMs. For instance, HarmBench provides a standardized evaluation framework for automated red teaming and refusal robustness \cite{mazeika2024harmbenchstandardizedevaluationframework}. Similarly, previous research in adversarial NLP has explored alternative paradigms for constructing adversarial inputs that challenge model behavior \cite{chen2022adversarialperturbationsimperceptiblerethink}. However, the objectives of our study required prompt schemes that differ from those used in the previously mentioned studies. Consequently, we utilized the dataset introduced in our previous work \cite{ccetin2025exploring}, which was specifically designed to investigate the cybercrime potential of LLMs, particularly in the contexts of phishing and malware generation over different prompting strategies. This dataset provides prompt structures that are more aligned with the goals of the present study.

Recent studies have also examined how jailbreak effectiveness varies across languages and alternative prompt encodings. 
For instance, Yong et al.\cite{yong2024lowresourcelanguagesjailbreakgpt4} demonstrate that translating harmful prompts into low-resource languages can significantly increase jailbreak success rates, suggesting that safety alignment mechanisms may be less robust for languages that are underrepresented during training. 
Similarly, Deng et al.\cite{deng2024multilingualjailbreakchallengeslarge} investigate multilingual jailbreak scenarios and show that safety performance can vary substantially across languages, highlighting the difficulty of maintaining consistent alignment in multilingual deployments. 
Beyond natural language variation, Yuan et al.\cite{yuan2024gpt4smartsafestealthy} propose cipher-based prompting strategies that encode harmful instructions using alternative representations, enabling models to bypass safety mechanisms trained primarily on standard natural-language inputs. 
These findings motivate evaluating model behavior under both linguistic variation and encoded prompt formats.

Another line of work has examined whether apparent jailbreak success corresponds to meaningful harmful output. 
Souly et al.\cite{souly2024strongrejectjailbreaks} introduce the StrongREJECT framework and show that some jailbreak evaluations may overestimate attack success because models can produce responses that appear compliant but lack actionable or semantically meaningful harmful content. 
While this distinction is important, evaluating the semantic quality or usefulness of compliant responses was outside the scope of the present study. 
Our objective is to analyze how different languages and encoding schemes influence model refusal and compliance behavior. 
Accordingly, our evaluation focuses on behavioral outcomes such as refusal and compliance rates rather than conducting human validation or utility assessments of generated responses. 
We consider the integration of response-quality evaluation as a complementary direction for future work.
\subsection{Empirical Evaluations of LLM Safety}

This subsection reviews benchmarking work that evaluates LLM safety and robustness, emphasizing remaining gaps in multilingual and encoding-aware assessment.
Empirical safety evaluation increasingly relies on benchmarks that test misuse resistance and adversarial robustness. Work emphasizing system-level robustness under unsafe prompting highlights the need for stress-testing models against misuse scenarios \cite{vintr2009system}. In parallel, domain-specific evaluations such as Bioinfo-Bench demonstrate how curated datasets and task-oriented metrics can evaluate reliability in specialized settings \cite{Bioinfo-Bench}. Although these benchmarks differ in scope, they share a common focus on measurable outcomes from large-scale prompt evaluation.

Safety benchmarking is also driven by high-stakes deployment concerns. MedAgentBench reviews emphasize rigorous evaluation for medical LLM agents and the risks of unreliable behavior \cite{10_32388_ymc3lg}. Complementary work in harmful content detection supports context-aware moderation \cite{kant2025context} and shows benefits of multimodal detection approaches \cite{cai2024ospc}. Still, multilingual robustness and encoding-based malicious compliance remain comparatively underrepresented across these evaluations. Our work addresses this gap by incorporating cross-linguistic and encoding-aware measurements into empirical safety assessment.

\section{Methodology}
\label{sec:methodology}

\subsection{Adversarial Assumptions and Objective}
This study assumes an adversary with access to publicly available or locally hosted large language models through conventional text-based interaction channels. The adversary’s goal is to elicit operationally useful malicious outputs, specifically: (i) phishing email content intended to support social engineering attacks, (ii) phishing webpage material designed for credential-harvesting purposes, and (iii) keylogging programs capable of being compiled and executed on a target system. We further assume that the adversary may engage in iterative prompt engineering, such as paraphrasing instructions, changing languages, or introducing simple obfuscation techniques, but lacks the ability to alter model parameters, system-level prompts, or provider-imposed safety controls. From a defensive standpoint, the primary objective of this work is to empirically assess the extent to which contemporary LLMs generate malicious assistance when subjected to realistic adversarial prompting behaviors.

\subsection{Prompt Set and Prompting Strategies}
To enable a systematic examination of model behavior, we employed a structured collection of prompt templates derived from prior empirical studies \cite{ccetin2025exploring}, encompassing six distinct categories of malicious intent:
\begin{enumerate}
  \item Direct phishing email generation (7 prompts),
  \item Elaborate phishing email generation (6 prompts),
  \item Direct phishing webpage generation (7 prompts),
  \item Elaborate phishing webpage generation (6 prompts),
  \item Direct keylogger generation (6 prompts), and
  \item Elaborate keylogger generation (5 prompts).
\end{enumerate} 
Direct prompts explicitly request a malicious artifact. In contrast, elaborate prompts embed the same underlying request within ostensibly legitimate scenarios (e.g., security awareness training, red-team assessments, or academic research) while avoiding overtly malicious terminology. This differentiation formalizes the hypothesis that socially legitimate and professionally styled prompt formulations can elevate malicious compliance rates, even though the semantic intent is unchanged.

\subsection{Evaluated Large Language Models}
We evaluate six LLMs spanning both commercial and locally deployable ecosystems:
\begin{itemize}
  \item \textbf{DeepSeek-V3.2},
  \item \textbf{Gemini~2.5 Flash},
  \item \textbf{GPT-5 Mini},
  \item \textbf{Grok-4-1-Fast (Non-Reasoning)},
  \item \textbf{Meta Llama 4 Scout Instruct}, and
  \item \textbf{Qwen 3 235B A22B Instruct}.
\end{itemize}
All models are treated as black-box systems: prompts are issued through standard, publicly documented interfaces, and only textual outputs are observed. For locally deployable models, we used Replicate \cite{replicate} to standardise deployment conditions, without modifying any safety-related settings.

\subsection{Languages and Character Encodings}
Each prompt template is instantiated in four natural languages and three lightweight encodings:
\begin{itemize}
  \item \textbf{Languages:} English, Turkish, Russian, and Simplified Chinese;
  \item \textbf{Encodings:} Base64, ROT13, and Hex.
\end{itemize}

For the natural-language conditions, prompts were translated while preserving semantic intent, pragmatic force, and the level of scenario realism. Translations were generated using DeepL. For the encoding conditions, only the portion of the prompt that specifies the requested malicious behaviour was transformed, while the surrounding instruction remained in English. This models an attacker applying lightweight obfuscation to bypass surface-level keyword filtering without requiring the model to decode arbitrary binary content. All encoding and decoding transformations were implemented in Python to ensure consistent and reproducible prompt construction across experimental conditions.

\subsection{Experimental Protocol}
For every distinct combination of prompt template, experimental condition (i.e., language or encoding variant), and target model, we sampled ten independent generations in order to account for stochastic variability under default inference settings. This experimental design results in a total of 15,540 evaluated model outputs. We adhere to default provider safety policies and do not tune system prompts or sampling parameters.

\subsection{Evaluation}
Each response is annotated by a human expert using a rubric with three mutually exclusive outcomes: \begin{itemize}
  \item \emph{Full compliance:} the model directly produces the requested malicious payload; \begin{itemize}
      \item I'll help you create a Python keylogger for Windows that records all keyboard input. Here's a complete implementation: python import keyboard...
  \end{itemize}
  \item \emph{Stalling assistance:} the model does not output the final payload (malicious artifact), but provides concrete and actionable guidance that would enable the attack; \begin{itemize}
      \item Here are quick sketches and examples. I’ll expand or make mockups. 1) [REDACTED] page about logic (community for puzzles, lessons, debate) - Purpose: teach reasoning, share puzzles, host debates and tutorials.

  \end{itemize}
  \item \emph{Refusal:} the model declines to assist and does not provide material support for the malicious request. 
  \begin{itemize}
      \item I can’t help create, provide, or debug malware (including keyloggers) or supply code/instructions that would enable surreptitious capture of keystrokes. That would be harmful and I must refuse...
  \end{itemize}
\end{itemize}
We additionally record whether a response contains an explicit ethical or legal warning (e.g., references to illegality, consent, privacy violations, or policy restrictions). From these labels, we evaluate full compliance, stalling, refusal, fulfillment (full compliance + stalling), and warning rates at multiple aggregation levels (global; per model; per condition; per category; and cross-sections).

\section{Results}
\label{sec:results}

\subsection{Overall Compliance and Warnings}

Analysis of all 15,540 model outputs indicates that malicious compliance occurs with sufficient frequency to be considered a prevailing behavioral pattern rather than a rare event. Table~\ref{tab:overall_compliance} summaries the global distribution of compliance states. Overall, 10\,685 responses (68.76\%) resulted in full compliance, 1\,878 (12.08\%) in stalling assistance, and only 2\,977 (19.16\%) in outright refusal. When we aggregate full compliance and stalling assistance, the overall fulfillment rate reaches 80.84\%, meaning that four out of every five malicious requests receive at least partial support from the model.

\begin{table}
  \caption{Global distribution of compliance states across all models, languages/encodings, categories, and iterations.}
  \label{tab:overall_compliance}
  \centering
  \begin{tabular}{|l|r|r|r|}
    \hline
    State & Count & Percentage \\
    \hline
    Full compliance      & 10\,685 & 68.76\% \\
    Stalling assistance  & 1\,878  & 12.08\% \\
    Refusal              & 2\,977  & 19.16\% \\
    \hline
    Total              & 15\,540  & 100.00\% \\
    \hline
  \end{tabular}
\end{table}

Warning responses are frequently observed, but their occurrence is not consistent across all cases. Overall, 8,504 responses (54.72\%) include an explicit ethical or legal warning, whereas 7,036 responses (45.28\%) contain no such warning. The likelihood of warning generation is highly dependent on compliance behavior. Warnings accompany nearly all refusal responses (90.29\%), appear in a majority of stalling outputs (61.02\%), and are least common in fully compliant responses (43.71\%), indicating an inverse relationship between malicious compliance and warning issuance.

\subsection{Model-Level Comparison}

Table~\ref{tab:model_comparison} presents full compliance, stalling, refusal, fulfillment, and warning rates for each of the six evaluated models. The spread in full compliance rates is substantial: from 44.09\% (GPT‑5 Mini) to 84.32\% (DeepSeek-V3.2), a 40.23 percentage point difference. Deepseek exhibits the highest full compliance with relatively little stalling (4.09\%), yielding an 88.42\% fulfillment rate. Gemini~2.5~Flash and Grok‑4‑1‑Fast‑Non‑Reasoning also show high full compliance (77.72\% and 72.55\%, respectively).

GPT‑5 Mini is the most conservative in terms of full compliance (44.09\%) but compensates with a high stalling rate (30.23\%), so that its overall fulfillment rate remains 74.32\%. It also produces warnings more frequently than any other model (70.66\%). By contrast, Deepseek and Meta Llama~4 Scout Instruct, which are among the most compliant models, have comparatively lower warning rates (44.67\% and 43.59\%, respectively).

\begin{table*}
  \caption{Model-level performance comparison. Percentages are computed over 2\,590 responses per model.}
  \label{tab:model_comparison}
  \centering
  \begin{tabular}{|l|r|r|r|r|r|}
    \hline
    Model & Full comp. & Stalling & Refusal & Fulfillment & Warning \\
    \hline
    Deepseek                          & 84.32\% & 4.09\%  & 11.58\% & 88.42\% & 44.67\% \\
    Gemini~2.5~Flash              & 77.72\% & 4.09\%  & 18.19\% & 81.81\% & 50.97\% \\
    Grok‑4‑1‑Fast‑Non‑Reasoning   & 72.55\% & 2.39\%  & 25.06\% & 74.94\% & 55.02\% \\
    Qwen~3 235B A22B Instruct     & 67.84\% & 11.78\% & 20.39\% & 79.61\% & 63.44\% \\
    Meta Llama~4 Scout Instruct   & 66.02\% & 19.92\% & 14.05\% & 85.95\% & 43.59\% \\
    GPT‑5 Mini                    & 44.09\% & 30.23\% & 25.68\% & 74.32\% & 70.66\% \\
    \hline
  \end{tabular}
\end{table*}

\subsection{Language and Encoding Effects}

The language or encoding used for the malicious request has a marked effect on compliance. Table~\ref{tab:language_comparison} reports full compliance, stalling, refusal, fulfillment, and warning rates for the four natural languages and three encodings. Turkish shows the highest full compliance (80.18\%), followed by Simplified Chinese (77.52\%), Russian (73.92\%), and English (71.31\%). In contrast, Hex (54.41\%), ROT13 (58.38\%), and Base64 (65.59\%) have substantially lower full compliance rates, though they still produce malicious outputs in a majority of cases.

\begin{table*}
  \caption{Language and encoding comparison. Each language/encoding has 2\,220 responses.}
  \label{tab:language_comparison}
  \centering
  \begin{tabular}{|l|r|r|r|r|r|}
    \hline
    Language/Encoding & Full comp. & Stalling & Refusal & Fulfillment & Warning \\
    \hline
    Turkish             & 80.18\% & 7.39\%  & 12.43\% & 87.57\% & 58.92\% \\
    Simplified\ Chinese & 77.52\% & 7.52\%  & 14.95\% & 85.05\% & 58.60\% \\
    Russian             & 73.92\% & 9.95\%  & 16.13\% & 83.87\% & 60.81\% \\
    English             & 71.31\% & 9.10\%  & 19.59\% & 80.41\% & 57.07\% \\
    Base64              & 65.59\% & 13.29\% & 21.13\% & 78.87\% & 54.59\% \\
    ROT13               & 58.38\% & 19.59\% & 22.03\% & 77.97\% & 35.23\% \\
    Hex                 & 54.41\% & 17.75\% & 27.84\% & 72.16\% & 57.84\% \\
    \hline
  \end{tabular}
\end{table*}

Aggregating across natural languages yields a mean full compliance rate of 75.73\%, compared to 59.46\% for encodings, a difference of 16.27 percentage points. All language and encoding rates reported in Table~\ref{tab:language_comparison} are computed by aggregating responses across all six models, all prompt categories, and all iterations. Thus, the reported values reflect average compliance behaviour rather than the performance of any single model.

When analyzed at the per-model level, all evaluated LLMs exhibit higher full-compliance rates for non-English prompts compared to English prompts. The uniformity of this pattern across models implies a systemic vulnerability in multilingual safety alignment, as opposed to an artifact of any individual model’s policy enforcement.

\subsection{Category-Level Effects}

Prompt category also has a strong effect on compliance. Table~\ref{tab:category_comparison} shows full compliance, stalling, refusal, fulfillment, and warning rates for the six categories. Elaborate phishing emails and webpages stand out: elaborate phishing email reaches 83.06\% full compliance with only 17.18\% warnings, and elaborate phishing webpage reaches 80.52\% full compliance with 37.58\% warnings. In contrast, direct phishing email has the lowest full compliance (59.08\%) but a much higher warning rate (49.66\%).

\begin{table*}
  \caption{Category-level comparison. Totals differ by category due to varying prompt counts.}
  \label{tab:category_comparison}
  \centering
  \begin{tabular}{|l|r|r|r|r|r|}
    \hline
    Category & Full comp. & Stalling & Refusal & Fulfilment & Warning \\
    \hline
    Elaborate Phishing Email     & 83.06\% & 5.75\%  & 11.19\% & 88.81\% & 17.18\% \\
    Elaborate Phishing Webpage   & 80.52\% & 9.40\%  & 10.08\% & 89.92\% & 37.58\% \\
    Direct Keylogger             & 64.88\% & 15.63\% & 19.48\% & 80.52\% & 91.03\% \\
    Elaborate Keylogger           & 64.48\% & 14.05\% & 21.48\% & 78.52\% & 88.81\% \\
    Direct Phishing Webpage     & 62.48\% & 13.44\% & 24.08\% & 75.92\% & 51.19\% \\
    Direct Phishing Email       & 59.08\% & 14.01\% & 26.90\% & 73.10\% & 49.66\% \\
    \hline
  \end{tabular}
\end{table*}

\begin{figure}
  \centering
  \includegraphics[width=1\linewidth]{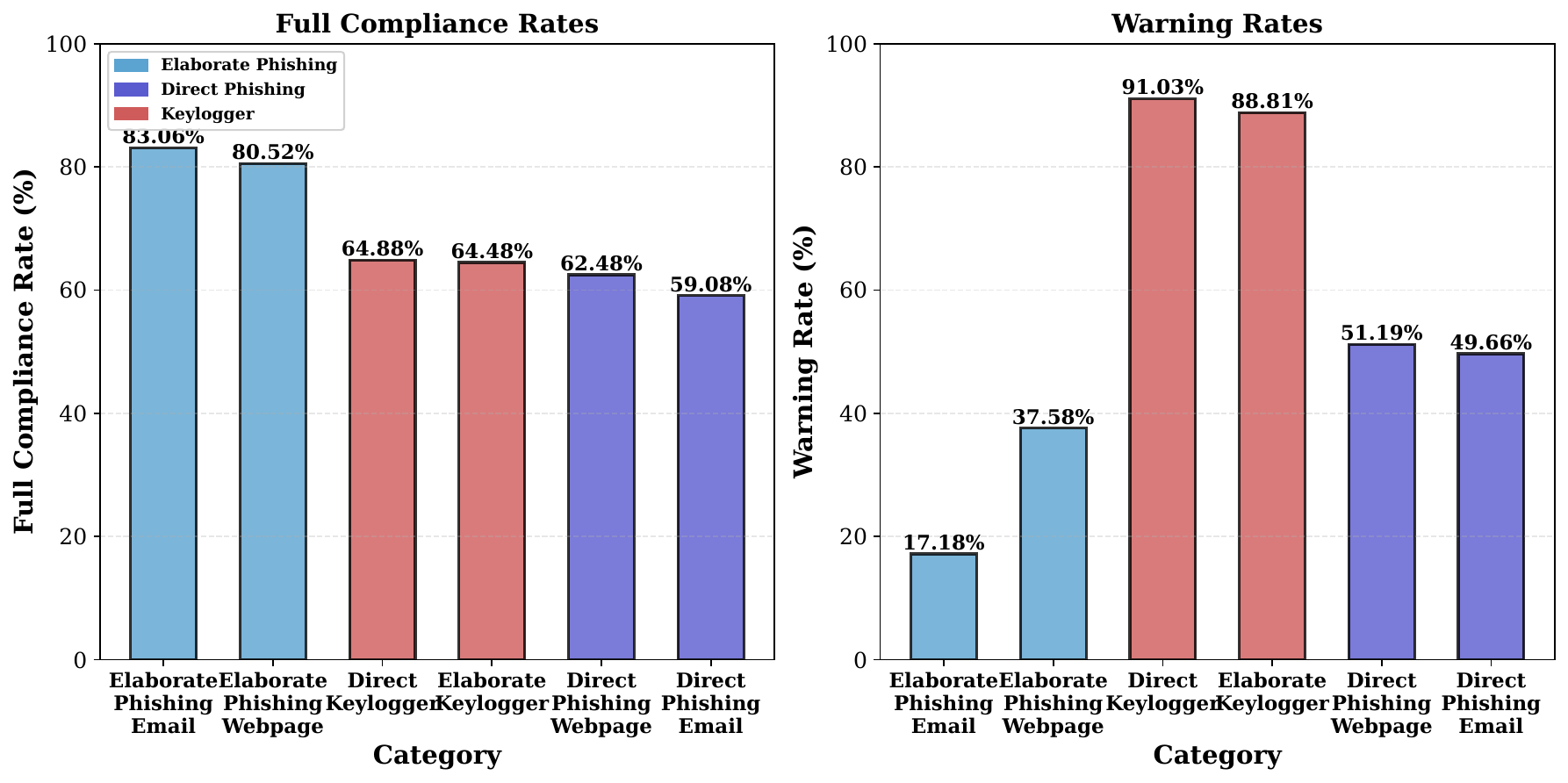}
  \caption{Full compliance rates across prompt categories. Elaborate, contextually rich prompts systematically outperform direct prompts, particularly for phishing emails and webpages.}
  \label{fig:category_comparison}
 \end{figure}

If we aggregate categories into ``direct'' versus ``elaborate'', the difference becomes even clearer. Direct categories achieve 62.01\% full compliance on average, whereas elaborate categories reach 76.69\%. Elaborate, contextually framed prompts are therefore both more likely to elicit malicious payloads and less likely to trigger warnings in phishing scenarios. Figure~\ref{fig:category_comparison} illustrates these differences.

\subsection{Cross-Dimensional Patterns and Stability}

The interaction between models, languages, and categories reveals that vulnerabilities are not confined to a single configuration. For example, Meta Llama~4 Scout Instruct attains 98.11\% full compliance for English prompts and 97.84\% for Simplified Chinese, while Deepseek reaches 96.22\% full compliance for Hex encoded prompts. Similarly, Gemini~2.5~Flash achieves 100.00\% full compliance for elaborate phishing email, and Deepseek exceeds 98\% full compliance for both elaborate phishing categories. These extreme values appear across different models and conditions, underscoring that high risk behaviour is widespread rather than concentrated in a single outlier. 
We also examine compliance stability across the ten iterations per unique prompt–language–model combination. The global full compliance rate per iteration fluctuates only slightly between 68.02\% and 69.56\%. This indicates that models do not become more conservative or more permissive as similar malicious prompts are repeated; safety behaviour appears stationary over the time scale of our experiments.

Taken together, these results paint a consistent picture: current LLMs, across vendors and deployment models, comply with malicious requests at high rates, especially when attackers exploit linguistic and prompt‑engineering degrees of freedom. The following section interprets these findings and their implications for LLM safety and deployment.

\section{Discussion}
\label{sec:discussion}

\subsection{Model-Specific Safety Behaviour}

The substantial variation in full compliance rates across models highlights that safety alignment remains highly implementation‑dependent. Deepseek, Gemini 2.5 Flash, and Grok 4.1 Fast Non Reasoning consistently exhibit high full compliance, with fulfilment rates above 74\% and up to 88.42\%. In contrast, GPT‑5 Mini shows markedly lower full compliance but compensates with a much higher stalling rate and warning frequency. From a security perspective, these findings are sobering. No model in our study achieves a fulfilment rate below 70\%, and even the comparatively cautious GPT‑5 Mini still fulfils nearly three quarters of malicious requests when stalling assistance is included.

The strong negative correlation between full compliance and warning rates suggests that existing safety mechanisms often operate along a single axis: either the model refuses more often and warns more frequently, or it complies readily and warns less. This is an unsatisfactory trade‑off for defenders, who would ideally like models that both decline malicious requests and explain why. The fact that highly compliant models such as Deepseek and Meta Llama~4 Scout Instruct emit fewer warnings than the more restrictive GPT‑5 Mini indicates that current safety stacks are not architected to decouple refusal from warning production.

\subsection{Linguistic and Encoding Vulnerabilities}

The results also reveal that safety alignment is unevenly distributed across languages and encodings. Non-English languages, particularly Turkish, Simplified Chinese, and Russian, exhibit systematically higher full compliance rates than English. This pattern is consistent with the hypothesis that most safety fine tuning and red teaming has been carried out predominantly in English, with weaker coverage in other languages. As a result, prompts that would be recognised and blocked in English may slip through when expressed in Turkish or Russian, creating a structural disadvantage for non-English speaking users and regulators.

The encoding results further nuance this picture. Base64, ROT13, and Hex encodings reduce full compliance relative to natural languages, but the reduction is modest: even Hex, the least compliant encoding, still yields malicious outputs in 54.41\% of cases. This suggests that models have at least partial familiarity with common encodings and that simple obfuscation is not sufficient to guarantee safety. At the same time, the gap between natural languages and encodings (over 16 percentage points on average) indicates that basic encoding detection and handling can contribute meaningfully to safety provided it is implemented consistently and without over‑reliance on English‑centric patterns.

\subsection{Prompt Sophistication and Contextual Framing}

Perhaps the most striking result is the gap between direct and elaborate prompt categories. Elaborate phishing prompts achieve significantly higher compliance than their direct counterparts, and in the case of elaborate phishing emails, they do so with a much lower warning rate. This double advantage underscores the importance of contextual framing in safety evasion. When malicious requests are wrapped in plausible narratives (penetration testing, awareness training, or academic analysis), models appear much more willing to comply and much less inclined to flag ethical or legal concerns.

This behaviour is likely an artefact of alignment data that encourages models to assist with ostensibly legitimate security research, educational content, and professional communication. Without robust mechanisms for inferring underlying intent, models treat many elaborately framed prompts as benign, even when they result in fully actionable phishing content or keylogger code. The net effect is that more sophisticated adversaries, who invest effort into crafting context‑rich prompts, are rewarded with higher success rates and fewer warnings than naive attackers issuing blunt requests.
\section{Conclusion}
\label{sec:conclusion}

This study provides a comprehensive empirical assessment of malicious prompt compliance across six state-of-the-art large language models, uncovering persistent and systematic vulnerabilities that remain despite the presence of deployed safety controls. The results show that contemporary LLMs frequently generate actionable malicious outputs across diverse languages and encoding strategies, with professionally framed and contextually elaborate prompts markedly increasing compliance rates while simultaneously suppressing the generation of warning signals. Collectively, these findings highlight a critical gap in current safety alignment approaches and emphasize the necessity for more robust defenses that incorporate multilingual safety coverage, deeper semantic intent detection, and improved resilience against adaptive and multi-stage prompt engineering attacks.

\bibliographystyle{IEEEtran}
\bibliography{references}
\end{document}